\documentclass[aps,prl,twocolumn,showpacs,superscriptaddress,groupedaddress]{revtex4-1}  
\usepackage{graphicx,color}  
\usepackage{dcolumn}   
\usepackage{bm}        
\usepackage{amssymb}   

\usepackage{amsmath} 
\usepackage{ulem}

\begin{document}
\title{From single-cell variability to population growth}
\author{Jie Lin, Ariel Amir}
\affiliation{Harvard John A. Paulson School of Engineering and Applied Sciences, Harvard University, Cambridge, Massachusetts 02138, USA}
\begin{abstract}
Single-cell experiments have revealed cell-to-cell variability in generation times and growth rates for genetically identical cells. Theoretical models relating the fluctuating generation times of single cells to the population growth rate are usually based on the assumption that the generation times of mother and daughter cells are uncorrelated. This assumption, however, is inconsistent with the exponential growth of cell volume in time observed for many cell types. Here we develop a more general and biologically relevant model in which cells grow exponentially and generation times are correlated in a manner which controls cell size. In addition to the fluctuating generation times, we also allow the single-cell growth rates to fluctuate and account for their correlations across the lineage tree. Surprisingly, we find that the population growth rate only depends on the distribution of single-cell growth rates and their correlations. 
\end{abstract}
\maketitle
\section{I. Introduction}
Even within a genetically identical population, phenotypes at the single-cell level can exhibit significant fluctuations \cite{Elowitz2002,Symmons2016}. A notable example is the fluctuation in the generation time (also called doubling time) \cite{Wang2010, Campos2014, Soifer2016,Hashimoto2016,Jafarpour2018,Eun2018}. Theoretical and experimental works suggest that the phenotypic heterogeneity in gene expression levels can enhance the population's fitness, {\it e.g.} through bet-hedging in a fluctuating environment \cite{Balaban2004, Kussell2005, Avery2006, Dhar2007, Raj2008, Ackermann2015a}. However, the effects of cell-to-cell variability on the population's fitness in a fixed environment have received much less attention. In conditions where cells proliferate with adequate resources, the number of cells increases with time as $N(t)\sim e^{\Lambda_p t}$. The population growth rate $\Lambda_p$ is often the dominant trait determining the fitness of the population \cite{Lenski1991,Vasi1994}. Early seminal work by Powell \cite{Powell1956} concluded that the fluctuation in generation times given a fixed mean increases the population growth rate under the key assumption that the generation times of mother and daughter cells are random and uncorrelated. This independent generation time assumption leads to analytically solvable models, but is also challenged by recent experimental observations showing for various microbial cells a finite correlation between mother and daughter cells' generation times \cite{Stewart2005,Taheri2015,Hashimoto2016,Wallden2016,Cerulus2016}. In fact, such correlations are inevitable given the exponential growth of cell volume at the single-cell level, acting as a feedback mechanism to control cell size \cite{Amir2014}.

For this reason, in this work we consider a more realistic model taking into account cell size control phenomenologically. Within Powell's model, the distribution of individual cell's generation times is the sole input to the model. However, in models in which cell size is controlled, there is one additional variable associated with each individual cell: its size must be explicitly specified as well. Note that as mentioned above the existence of size control must causally lead to correlations among, e.g.,  mother and daughter generation times, which are indeed observed. Nevertheless, the converse is not true, and considering a model in which cell size is not explicitly included but in which generation times are correlated -- will not control cell size (since there is no way to correct for fluctuations in cell size without having explicit information about size). This fundamental point precludes many of the previous works on the subject from being biologically realistic, even works in which generation times are correlated \cite{Cerulus2016}. In order to control it, size must be accounted for explicitly as a model variable, in addition to the generation time.

The structure of the manuscript is as follows. We first describe our model and define the phenomenological size control strategy, building on previous works, which dictates the structure of the correlations on the lineage tree. Next, we define a biologically inadequate model, in which generation times are correlated between mother and daughter cells but cell size is not considered. This model has also been defined (but not solved) by Powell in his seminal work \cite{Powell1956}. Using a novel approach, we are able to provide an analytic solution to this simpler model. This solution serves as a stepping stone for attaining an analytical solution for the biologically relevant model discussed above. We find that, intriguingly, variability associated with the timing of divisions (without affecting the single-cell growth rate) has negligible consequence on the population growth rate. In contrast, the statistics of the single-cell growth rate variability (both its magnitude and the strength of its correlation across the lineage tree) affect the population growth rate. We find that the single-cell growth rate variability can either increase or decrease the population's growth rate, with a Pearson correlation coefficient approximately equal to $1/2$ separating the two regimes. It appears that many microbial populations support a growth rate correlation weaker than this threshold value \cite{Stewart2005,Logsdon2017,Eun2018}, suggesting that variability will typically be detrimental to the population growth.

\begin{figure}[htb!]
	\includegraphics[width=.45\textwidth]{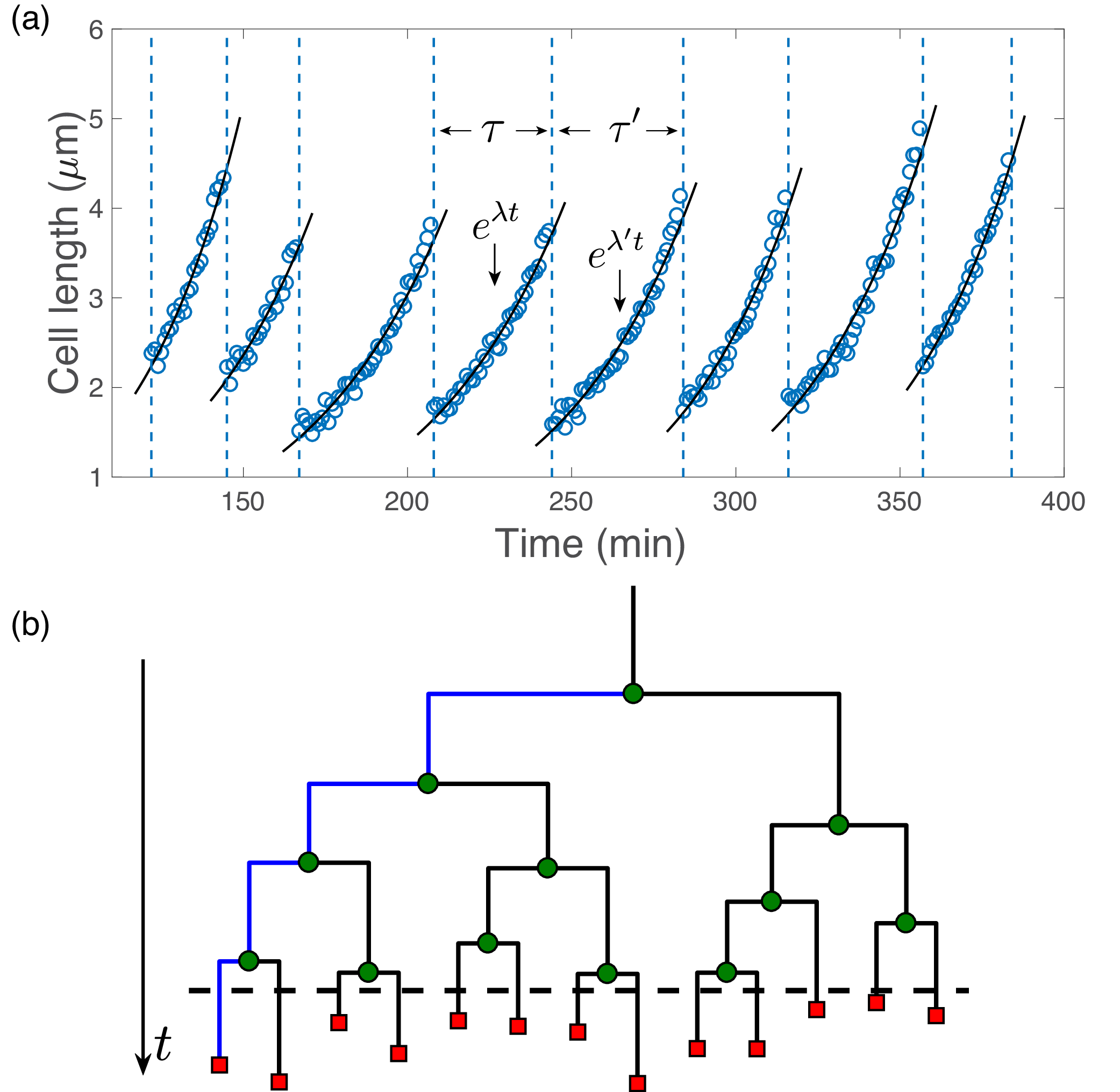}
	\caption{(a) Cell length {\it v.s.} time along a single lineage. The black lines are the exponential fits of cell length in time. The dashed lines mark the boundaries between consecutive generations. Data from Ref .\cite{Tanouchi2017}. (b) A lineage tree starting from a single cell. After a long enough time, the number of cells increases as $N(t)\sim e^{\Lambda_p t}$ in time.}\label{figure1}
\end{figure}

\section{II. Size-controlled model} 
Recent work on microbial growth across the three domains of life has shown that it can adequately be described phenomenologically as as discrete map (i.e., an autoregressive stochastic process), whereby phenotypes in the next generation are related to those in the previous generation alone. Furthermore, cell size growth appears to be well approximated by an exponential growth in time within each cell cycle, with the single-cell growth-rate a fluctuating variable that is potentially correlated between mother and daughter cells (Fig. \ref{figure1}(a)). These insights can be summarized succinctly in the following three equations, which form the heart of the model we use throughout the paper:
\begin{align}
&\ln(\lambda^{\prime})=C_{\lambda}\ln(\lambda)+B_\lambda+\eta, \label{model_eq_1}\\
&V_d=e^{\lambda\xi}\big(2(1-\alpha)V_b+2\alpha V_0 +2 \delta\big),\label{model_eq_2}\\ 
&\tau=\frac{1}{\lambda}\ln(\frac{V_d}{V_b}).\label{model_eq_3}
\end{align}
The first equation is an autoregressive model for the logarithm of the single-cell growth-rate, $\lambda$. This ensures that the growth rate remains strictly positive. The variable $C_{\lambda}$ can be readily shown to be the Pearson correlation coefficient between $\ln(\lambda)$ of mother and daughter cells (and for small noise, approximately equal to the growth rate correlation). The parameter $B_{\lambda}$ is constant, while $\eta$ is a noise term assumed to be normally distributed with zero mean and variance $\sigma_{\eta}^2$. Similarly, the second equation is an autoregressive model for the cell size. The cell size strategy, which has received significant attention in recent studies, is governed by the parameter $\alpha$ (where, e.g., $\alpha=1$ corresponds to a critical cell size for division and $\alpha=1/2$ to a constant addition of volume from birth to division). Note that here $\delta$ and $\xi$ are respectively the size-additive and time-additive noise. Both of them are normally distributed with zero mean and variances equal to $\sigma_{\delta}^2$ and $\sigma_{\xi}^2$ respectively. The details of the choice of noise statistics will not be important to the results discussed here, as we show later. Finally, the third equation relates the generation time to the cell size at birth and division and the growth-rate in that generation. We will refer to the above model as the size-controlled model (SCM). We note that to fully specify the population growth on a lineage tree, we also need to specify the sister-sister correlations. As proven in Ref. \cite{Powell1956}, the sister-sister correlation does not affect the population growth rate (this also follows from our recursive approach as we show in Appendix B). For this reason, we simplify the model by assuming $\delta$ and $\xi$ are independent between sister cells

We use the model equations to ``grow" a lineage tree, such as that illustrated in Fig. \ref{figure1}(b). Such a lineage tree will have correlated logarithmic growth-rates (with mother-daughter Pearson correlation coefficient of $C_\lambda$), correlated cell size (with mother-daughter Pearson correlation coefficient of $1-\alpha$), and correlated generation times, which are contributed both by the cell size control and the correlated growth rates. The number of cells in this branching processes will scale as $N(t) \propto e^{\Lambda_p t}$. The quantity $\Lambda_p$ defines the proliferation of the population, and is of central importance to characterizing the population fitness. Although simple to define and simulate numerically, our main goal in this work is to understand how $\Lambda_p$ depends on the parameters $C_{\lambda}$, $\alpha$ and the magnitude of the noise terms. Given a lineage tree, one may consider the distribution of relevant phenotypes along a single lineage, which is denoted as the lineage distribution of that particular phenotype (blue lines in Fig.\ref{figure1}b). Other types of distributions can also be defined \cite{Lin2017}, {\it e.g.}, the tree distribution, which is based on the statistics of all cells in the tree including the ``branch cells" which have already divided (green circles) and the ``leaf cells" which are currently present (red squares), see Fig. \ref{figure1}(b).

\section{III. A digression: Random-generation-time model}
Consider a different model, in which the logarithm of generation times of mother ($\tau$) and daughter cells ($\tau^{\prime}$) are coupled through an autoregressive process \cite{Amir2014, Ho2018},
\begin{equation}
\ln(\tau^{\prime})=C_{\tau}\ln(\tau)+ B_{\tau}+\kappa.\label{Langevin}
\end{equation}
Here $C_{\tau}$, $B_{\tau}$ are constant, and $\kappa$ is a normally distributed noise with zero mean and variance $\sigma_{\kappa}^2$. Eq. (\ref{Langevin}) defines the statistics along a single lineage. As we show in Appendix B and noted above, the correlation in the noise term $\kappa$ between sister cells has no consequences for the population growth. 
Note that even when $\kappa$ between sister cells are not correlated, the Pearson correlation coefficient between siblings will not vanish, since they are correlated via the mother cell. In this case, the Pearson correlation coefficient between the logarithmic generation times of mother and daughter cells is $C_{\tau}$ and the sister-sister correlation coefficient is $C_{\tau}^2$.

As noted above, this class of models is not equivalent to the one defined earlier, since although the presence of size control necessarily implies the existence of correlations in the generation time, the converse is not true, and the correlations in generation times cannot correct the size fluctuations appropriately given that cell growth is exponential in time. Nonetheless, precisely this model has been proposed already in the 1950's in Powell's seminal work \cite{Powell1956}, and variants of it has been utilized also in recent works, in the context of the population growth of asymmetrically dividing budding yeast \cite{Cerulus2016}. In fact, the random-generation-time model is equivalent to the ``kicked cell cycle" model in Refs. \cite{Sandler2015,Mosheiff2018} without the external forcing term due to circadian clocks. Intriguingly, it has been found that the circadian clocks can break down the simple relation between the sister-sister correlations and mother-daughter correlation and similar effects can also be found in size-controlled model \cite{Lin2017}. We are able to provide an analytic solution to this model, which will mathematically serve us in solving the more biologically realistic size-controlled model.

We provide the detailed derivations of the population growth rate in Appendix B and show the final formula here:
\begin{equation}
\Lambda_p=\frac{2\ln(2)/\langle \tau \rangle}{1+\sqrt{1-2\ln(2)\sigma_{\tau}^2/\langle \tau \rangle^2 F(C_{\tau})}},\label{lambda_p}
\end{equation}
with $F(C_{\tau})=(1+C_{\tau})/(1-C_{\tau})$, $\langle\tau\rangle=\exp( \frac{B_{\tau}}{1-C_{\tau}}+\frac{\sigma_{\kappa}^2}{2(1-C_{\tau}^2)})$ and $\sigma_{\tau}^2=\langle\tau\rangle^2(\exp(\frac{\sigma_{\kappa}^2}{1-C_{\tau}^2})-1)$ as the mean and variance of generation times along single lineages. This shows that increasing the variance in the generation time -- for a fixed mean generation time -- leads to a larger population growth rate, in line with Powell's conclusions from studying the uncorrelated case (corresponding to $C_{\tau}=0$). Furthermore, this equation implies that the population growth rate monotonically increases as the correlation coefficient $C_{\tau}$ increases, consistent with previous simulations \cite{Cerulus2016}. This can be intuited by noting that in this scenario once a cell with a short generation time occurs randomly its offsprings will also have a shorter generation time and will contribute significantly to the population growth. It is true that the same holds for cells with a large doubling time, but since the number of cells grows exponentially, the population gains more from the former scenario than it loses from the latter.

\begin{figure}[htb!]
   \includegraphics[width=.4\textwidth]{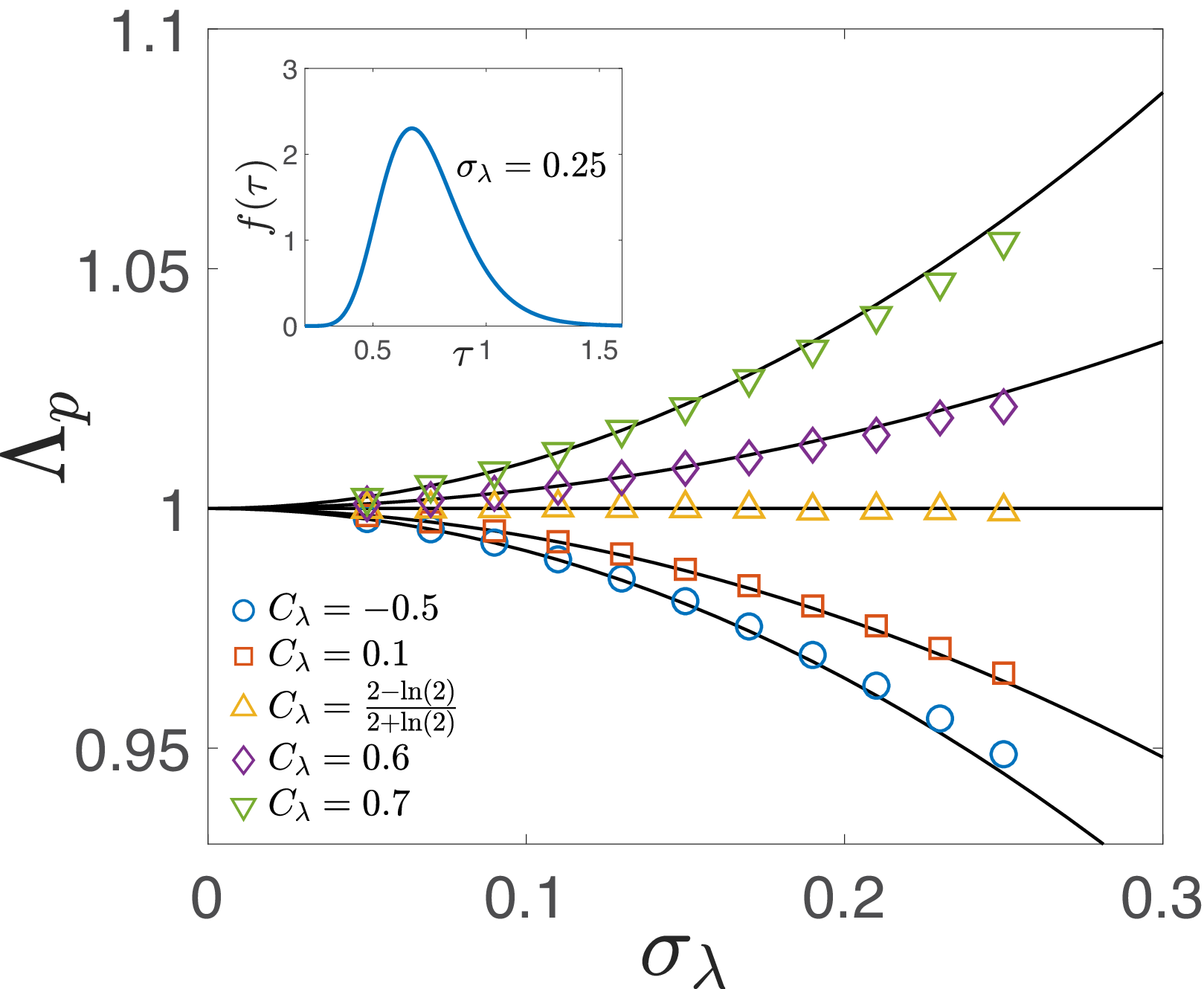} 
    \caption{In this figure, the variability in generation times arises only from the variability in single-cell growth rates. The figure shows $\Lambda_p$ {\it v.s.} $\sigma_{\lambda}$, compared with the theoretical prediction of Eq. (\ref{lambda1}) (black lines). The correlation coefficient $C_{\lambda}$ is indicated in the legend. The inset shows the lineage distribution of generation times with $\sigma_{\lambda} = 0.25$, which is non-Gaussian.}\label{figure3}
\end{figure}

\section{IV. Random-growth-rate model} As discussed above, the random-generation-time model is not biologically adequate since cell size control is unaccounted for. In the following, we study a scenario in which cell size is controlled and the generation time is set by the exponential growth rate of cell volume -- which in this model will be a stochastic variable. In the simplest scenario (which we will relax in the next section) we assume the cell size is perfectly regulated so that each cell divides at the cell volume $V_d=2$ symmetrically. Perfect size regulation corresponds to the limit where $\xi=\delta=0$ in Eq. (\ref{model_eq_2}), and the cell volumes at birth and division will quickly converge to $V_b=V_0\equiv 1$ and $V_d=2V_0\equiv 2$ independent of the initial conditions. Therefore the generation time equals:
\begin{equation}
    \tau=\ln(2)/\lambda,\label{time_growth}
\end{equation}
where $\lambda$ is the single-cell growth rate. Since the variability in generation times arises exclusively from the variability in growth rates, we refer to this model as the random-growth-rate model.

The autoregressive model of Eq. (\ref{model_eq_1}) would generally generate a log-normal growth rate distribution \cite{Amir2014}:
\begin{equation}
    P(\lambda) = \frac{1}{\lambda \sigma \sqrt{2 \pi}}\exp\Big(-\frac{(\ln(\lambda)-\mu)^2}{2\sigma^2}\Big),\label{lambda_time}
\end{equation}
with $\mu =B_\lambda/(1-C_\lambda)$ and $\sigma^2 =\sigma^2_{\eta}/(1-C^2_\lambda) $. 
Using Eq. (\ref{lambda_time}) we can readily calculate the form of the generation time distribution, finding that its mean and variance are:
\begin{equation}
   \langle \tau \rangle = \ln(2) e^{\frac{\sigma ^2}{2}-\mu }, \quad \sigma^2_\tau = \langle \tau \rangle^2 (e^{\sigma^2}-1). \label{mean_time}
\end{equation}
For small $\sigma_{\lambda}$ the mother-daughter Pearson correlation coefficient of generation times approximately equals that of the growth rates: since for small noise both generation time and growth rates will manifest small fluctuations around the steady-state value, we may Taylor expand Eq. (\ref{time_growth}) around the mean generation time, and utilize the fact that Pearson correlation coefficients are insensitive to shifts and scaling \cite{Amir2014}. For this reason we may utilize the results of the random-generation-time model Eq. (\ref{lambda_p}), using the above values of mean and variance for the generation time, and setting $C_{\tau}=C_\lambda$. To the lowest order of $\sigma_{\lambda}/\langle \lambda\rangle$, we obtain:
\begin{equation}
\Lambda_p/\langle \lambda \rangle \approx 1-\Big(1-\frac{\ln(2)}{2}\frac{1+C_{\lambda}}{1-C_{\lambda}} \Big)\sigma_{\lambda}^2/{\langle \lambda \rangle}^2.\label{lambda1}
\end{equation}
From the above equation, we find a critical correlation coefficient of single-cell growth rates, $C_{\lambda}=\frac{2-\ln(2)}{2+\ln(2)}\approx 0.5$, separating two scenarios: if $C_{\lambda}< 0.5$, the variability in single-cell growth rates decreases the population growth rate; in contrast, if $C_{\lambda}>0.5$, the variability in single-cell growth rates increases the population growth rate. This result is consistent with our previous numerical finding \cite{Lin2017}. We compare the numerical results of $\Lambda_p$ with Eq. (\ref{lambda1}) in Fig. \ref{figure3}, finding satisfying agreement for small $\sigma_{\lambda}$. Interestingly, the theory provides an excellent approximation to the population growth rate even in scenarios where the generation time distribution manifests significant deviations from a normal distribution (inset of Fig. \ref{figure3})

\begin{figure}[htb!]
   \includegraphics[width=.45\textwidth]{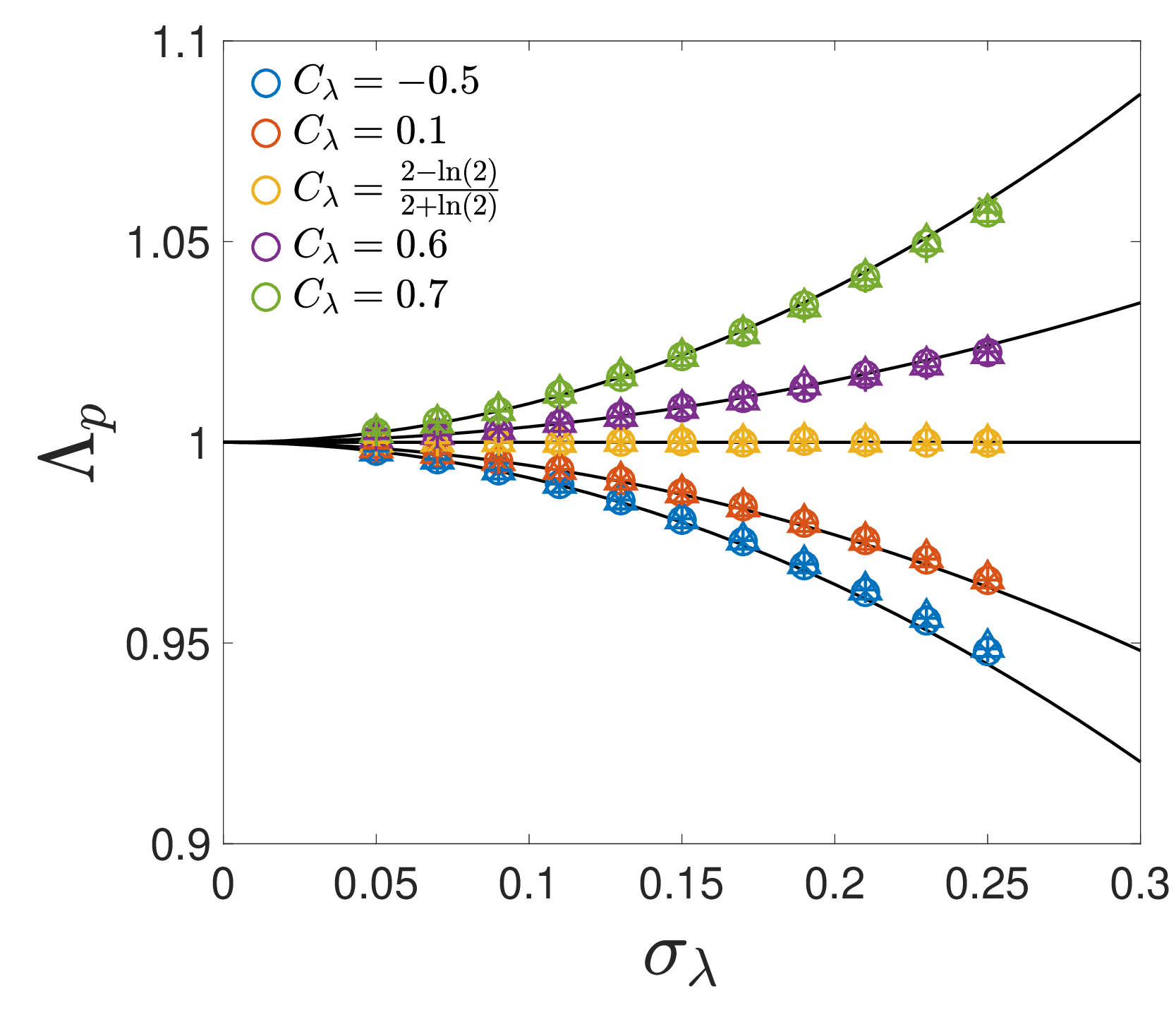} 
    \caption{The dependence of the population growth rate on the growth rate noise is well captured by Eq. (\ref{lambda1}) (black lines) even in the presence of other sources of noise. Colors represent different $C_{\lambda}$ as shown in the legend. Both simulations with time-additive noise ( $\sigma_{\xi}=0.1$ with $\alpha=0.5$ ($\bigcirc$) and $\alpha=1$ ($\bigtriangleup$)) and size-additive noise ( $\sigma_{\delta}=0.1$ with $\alpha=0.5$ ($\times$) and $\alpha=1$ ($+$)) are shown.}\label{figure4}
\end{figure}

\section{V. Towards a biologically realistic population growth model} 
The two previously discussed models are both biologically unrealistic. The random-generation-time model is unreasonable since it does not account for cell size, which we know to be under tight control for all organisms in nature. The random-growth-rate model assumes perfect size control, and does not account for fluctuations in cell size. For these reasons we now turn back to the original model described by Eqs. (\ref{model_eq_1})-(\ref{model_eq_3}), which previous work has has suggested to be an adequate 	phenomenological description of growth in microbes \cite{Ho2018}. In principle, the population growth rate in the full model may depend on the size-control-strategy parameter $\alpha$, the variance of the size-additive ($\sigma_{\delta}^2$) or time-additive noise ($\sigma_{\xi}^2$), the relative growth-rate fluctuations ($\sigma_{\lambda}^2/\langle \lambda\rangle^2$), and the growth-rate correlations controlled by $C_\lambda$. However, a simple argument hints that out of these multiple parameters, the only relevant ones are those related to the growth-rate statistics ($\sigma_{\lambda}^2/\langle\lambda\rangle^2$ and $C_\lambda$).

To see this, we first note that when the single-cell growth rate does not fluctuate ($\sigma_{\lambda} =0$) and equals $\lambda_0$, then $\Lambda_p = \lambda_0$ precisely independent of the size-control-strategy parameter $\alpha$, time-additive and size-additive noise \cite{Lin2017}. This enormous simplification comes about since in the presence of size-control the entire volume of the population must grow exponentially at the same rate as the number of cell. The former quantity, however, is agnostic of the timing of cell divisions -- as long as cell growth is exponential at the single-cell level, the separation of one cell into two does not affect the rate of population volume accumulation.

Assuming a general form of $\Lambda_p(\sigma_{\lambda}^2/\langle\lambda\rangle^2,\sigma_\xi^2,\sigma_{\delta}^2)$ that depends on the variance of each of the difference stochastic contributions, we may Taylor expand it and conclude that the leading order terms cannot contain terms that do not couple to the growth-rate variance (since when it is zero, $\Lambda_p$ is constant). Therefore the population growth rate must take the following form to leading order:
\begin{equation}
\Lambda_p/\langle \lambda \rangle \approx 1 + f(\alpha,C_\lambda)\sigma_\lambda^2/\langle \lambda \rangle^2 .\label{Taylor}
\end{equation}
In the limit $\sigma_{\xi}=\sigma_{\delta}=0$, cell size is perfectly regulated and independent of $\alpha$ according to Eq. (\ref{lambda1}). Therefore, we conclude that $f$ is independent of $\alpha$ and depends only on $C_\lambda$.

This argument suggest that the results of Eq. (\ref{lambda1}) should also provide a good approximation to the complete scenario described by Eqs. (\ref{model_eq_1})-(\ref{model_eq_3}), where the model also includes size-additive and time-additive noise that affect both the size distributions and the generation time distributions (but importantly, not the single-cell growth rate statistics). 
We verify this in Fig \ref{figure4}, showing that Eq. (\ref{lambda1}) (black lines) provides a good approximation to the numerical values of $\Lambda_p$ independent of the size-control parameter $\alpha$, the size-additive or time-additive noise. Interestingly, the theory provides an excellent approximation to the population growth rate even though the original prediction of random-generation-time model (Eq. (\ref{lambda_p})) no longer works (Fig. 5(b)). We also discuss the higher-order corrections due to the size-additive and time-additive noise in Appendix E.

\section{VI. Discussion} 
In this work, we study the growth of an exponentially proliferating population of cells with phenotypic variability that is correlated across the lineage tree. We first solve the long-standing problem of the random-generation-time model, first proposed by Powell in 1956. The results indicate that a positive correlation between mother and daughter cells' generation times increases the population growth rate given a fixed mean generation time. Recent works in the same spirit of Powell \cite{Cerulus2016,Jafarpour2018} have extended the random-generation-time model to asymmetric division and obtained similar conclusions. However, as Powell's model, these models are not biologically realistic for microbes since it does not account for cell size control and lead to incorrect prediction as we show in Fig. 5(b). Nevertheless, we were able to utilize the exact solution of this model to a biologically realistic scenario in which the cell volume grows exponentially and cell size is controlled. Compared with previous works, we captured the relevant single-cell phenotype that affects the population growth rate and found that the single-cell growth rate variability lowers the population growth rate when the growth rate correlation is lower than a threshold value of about $1/2$. For strongly correlated growth rates, variability enhances the population growth. Interestingly, within this model the cell size control strategy has no effect on the population growth, and forms of stochasticity that affect the generation time and size but not the single-cell growth rate itself -- have no consequence for the population growth rate. 

Under the assumption that the model described by Eqs. (1)-(3) is valid, our results provide a recipe to infer the population growth rate based on single-lineage data, {\it e.g.}, data from mother-machine experiments \cite{Wang2010}. All one needs to know is the average, the standard deviation and the mother-daughter correlation of single-cell growth rates according to Eq. (\ref{lambda1}). For instance, we use the lineage data from Ref. \cite{Tanouchi2017} to compute the population growth rate and obtain $\Lambda_p=1.51 h^{-1}$ (Fig. \ref{figure1}(a)), which is distinct from the prediction according to the random-generation-time model, $\Lambda_p=1.33h^{-1}$.

Our results have an intriguing evolutionary implication. For those organisms with a strong correlation in growth rates between mother and daughter cells, heterogeneity in growth rates is evolutionary favorable, while for organisms with a weak correlation between growth rates, evolutionary selection would tend to minimize the heterogeneity in growth rates. The correlation coefficients of single-cell growth rates have been reported for {\it E. coli} and a wide range of values have been reported from close to $0$ to $0.7$ \cite{Stewart2005, Lin2017, Kiviet2014,Taheri2015,Wallden2016,Kennard2016}, which establishes the relevance of our study to microbial evolutionary dynamics. Since the cell volume of cancer cells also grows exponentially \cite{Cermak2016}, our results may shed light on the evolutionary dynamics of cancer cells as well.

\begin{acknowledgments}
 We thank Felix Barber, Hanrong Chen, Po-Yi Ho, Ethan Levien and Felix Wong for useful discussions. AA thanks the A.P. Sloan foundation, NSF CAREER 1752024, the Volkswagen Foundation and Harvard Dean's Competitive Fund for Promising Scholarship for their support. JL was supported by the George F. Carrier Fellowship of Harvard's SEAS.
 \end{acknowledgments}

\renewcommand{\theequation}{A\arabic{equation}}
\setcounter{equation}{0}

\section{Appendix A: Numerical simulation details}
We simulate an asynchronous population and compute the resulting population growth rate. For the random-generation-time model, the simulation starts from $100$ cells with random generation times sampled from a lognormal distribution with mean $1$ and variance $\sigma_{\tau}^2$ and uniformly distributed relative ages from $0$ to $1$.  For the random-growth-rate model and size-controlled-model, the simulation starts from a single cell with a random growth rate and the logarithm of its growth rate is sampled from a Gaussian distribution with zero mean and variance $\sigma_{\eta}^2/(1-C_{\lambda}^2)$. The population growth rate in the exponentially growing phase is independent of the initial conditions. We run the simulation until there are $N=5\times 10^{6}$ number of cells, and compute the population growth rate using the data in the final window with a time interval $\Delta t=5$.

\begin{figure}[htb!]
	\includegraphics[width=.4\textwidth]{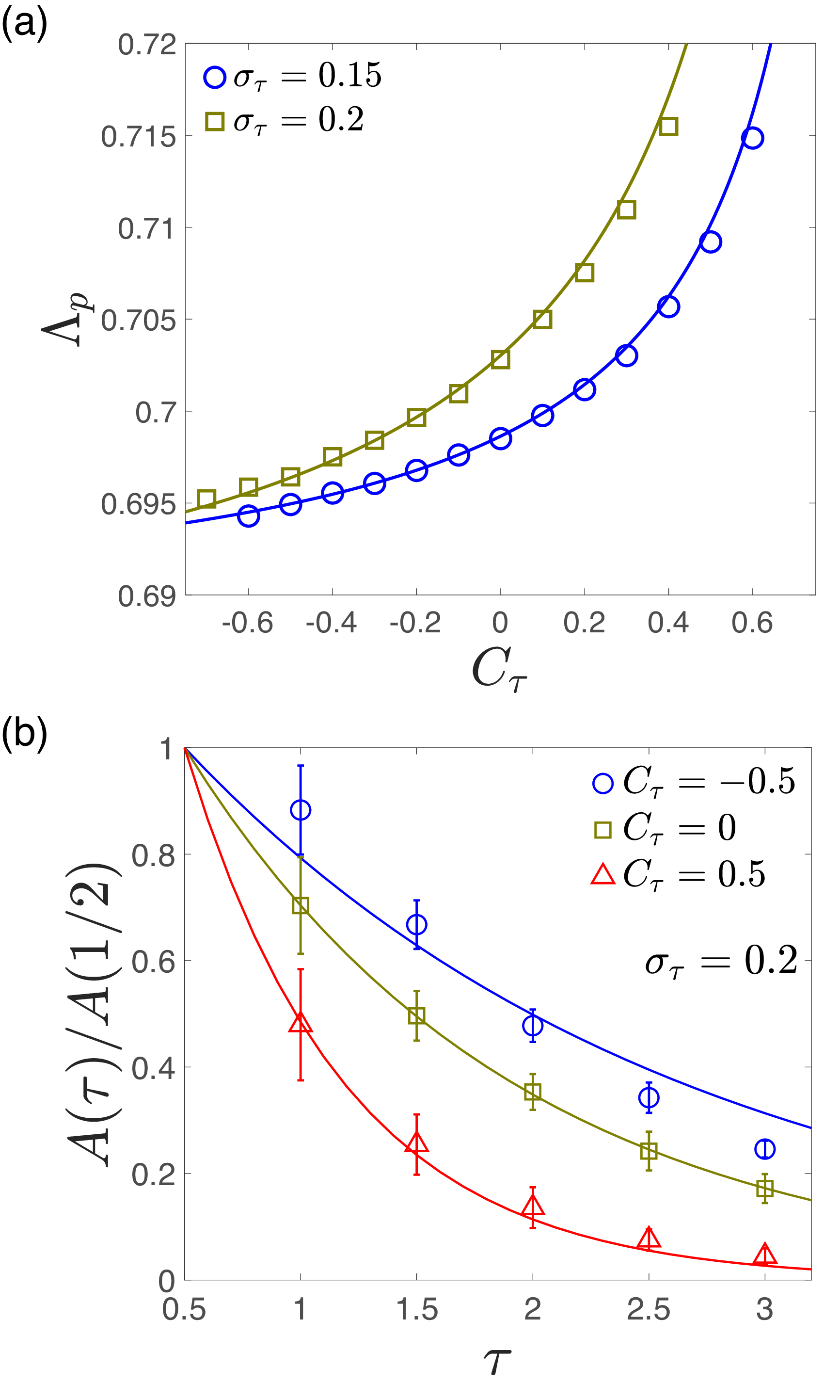} 
	\caption{(a) Numerical simulations of the population growth based on the random-generation-time model. $C_{\tau}$ is the correlation coefficient between the generation times of mother and daughter cells. $\sigma_{\tau}$ is the standard deviation of generation times. The solids lines are the theoretical predictions, Eq. (5) in the main text. Here, the mean generation time is fixed to be $1$. (b)The amplitude of the exponential growth of the population {\it v.s.} the generation time of the first cell. The error bar is the standard deviation of $30$ samples. }\label{figures1}
\end{figure}

\begin{figure}[htb!]
	\includegraphics[width=.4\textwidth]{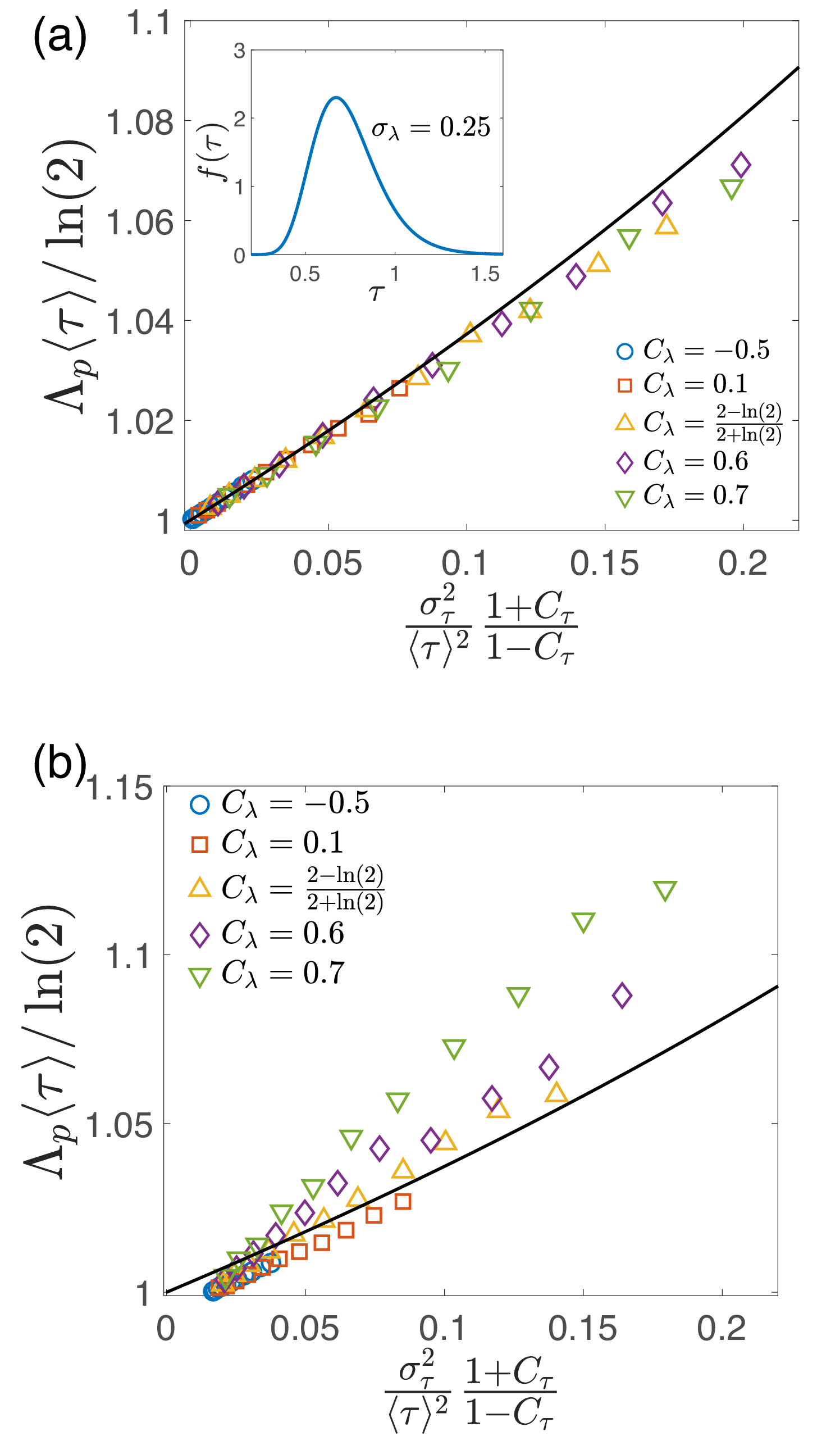} 
	\caption{(a) The normalized population growth rate $\Lambda_p\langle \tau\rangle/\ln(2)$ {\it v.s.} $\sigma_{\tau}^2/\langle\tau\rangle^2 \frac{1+C_{\tau}}{1-C_{\tau}}$ for the random-growth-rate model. Each symbol represents a different $C_{\lambda}$ with changing $\sigma_{\lambda}$. All the data collapse on the solid line, which is the theoretical prediction from the random-generation-time model, Eq. (5) in the main text. The inset shows the lineage distribution of generation times at $\sigma_{\lambda}=0.25$, which is non-Gaussian. (b) The same analysis for the size-controlled model where the prediction of the random-generation-time model breaks down. Here $\sigma_{\delta}=0$, $\sigma_{\xi}=0.1$. }\label{figures2}
\end{figure}

\begin{figure}[htb!]
	\includegraphics[width=.4\textwidth]{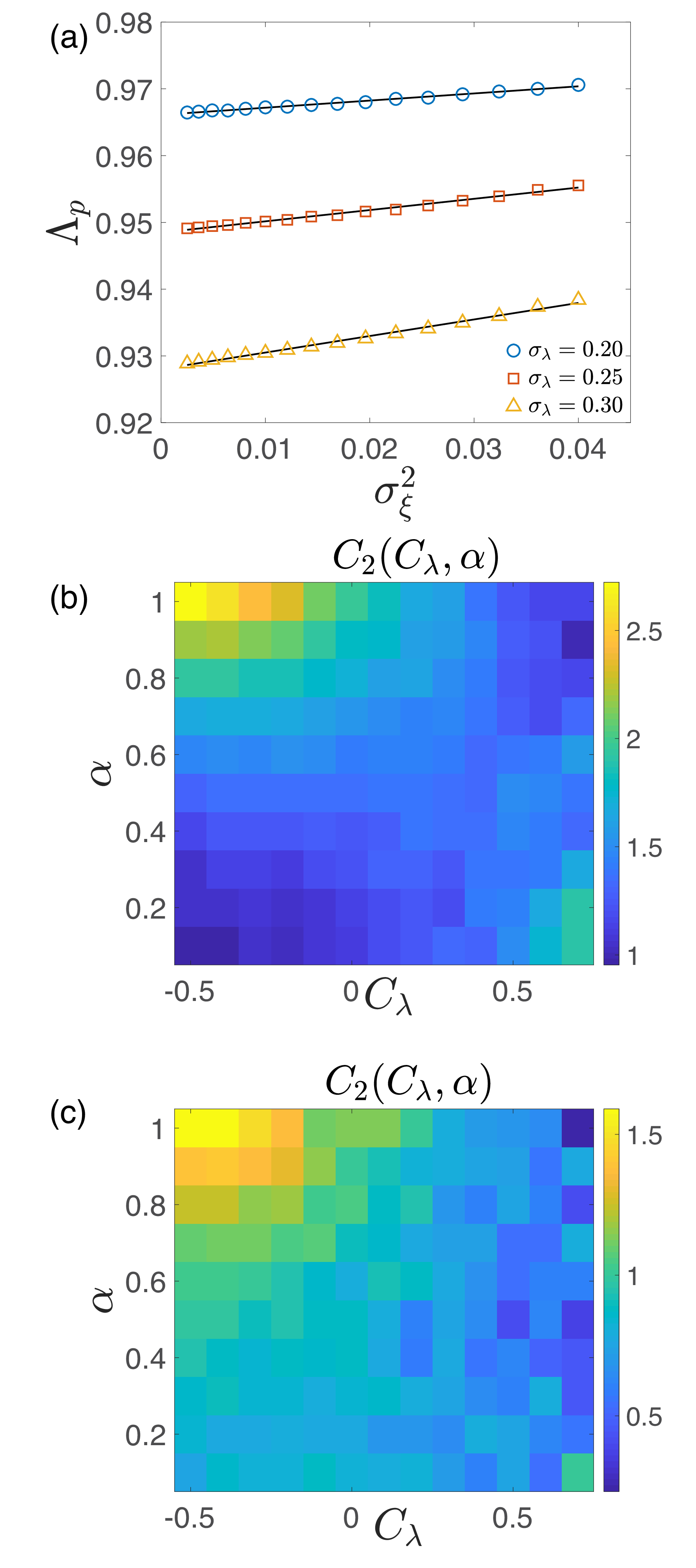} 
	\caption{(a) $\Lambda_p$ {\it v.s.} the variance of time additive noise, $\sigma_{\xi}^2$. The single-cell growth rate variabilities are indicated in the legend. The black lines are linear fittings based on Eq. (\ref{lambdap}). Here $C_{\lambda}=-0.5$, $\alpha=1$. (b) The fitted $C_2$ coefficients as functions of $C_{\lambda}$ and $\alpha$ for time-additive noise. (c) The fitted $C_2$ coefficients as functions of $C_{\lambda}$ and $\alpha$ for size-additive noise. }\label{figures3}
\end{figure}
\section{Appendix B: Derivation of the population growth rate of the random-generation-time model}
In the long time limit, the total number of cells grows exponentially with a constant population growth rate independent of the initial condition and the initial condition determines the transient dynamics before reaching the exponential growing steady state. We can define the amplitude of the growing population in the steady state as 
\begin{equation}
N(t)=A(\tau)\exp(\Lambda_pt),
\end{equation}
where $A(\tau)$ depends on the generation time of the first ancestral cell, $\tau$, see Fig. \ref{figure1}(b). The probability distribution of the generation times of the daughter cell is conditioned on the mother cell's generation times, $h(\tau^{\prime}|\tau)$. Using the self-similarity of the population tree, we obtain the recursive equation,
\begin{equation}
A(\tau)=2\int_0^{\infty} e^{-\Lambda_p \tau} A(\tau^{\prime}) h(\tau^{\prime} |\tau)d\tau^{\prime}.\nonumber
\end{equation}
Given $h(\tau^{\prime}|\tau)$, there appear to exist a unique set of $\Lambda_p$ and $A(\tau)$ (up to a multiplicative factor for $A(\tau)$) that satisfy the above equation. To further simplify the problem, we define $B(\tau)=A(\tau)\exp(\Lambda_p \tau)$ and obtain
\begin{equation}
B(\tau)=2\int_0^{\infty} e^{-\Lambda_p \tau^{\prime}} h(\tau^{\prime} |\tau)B(\tau^{\prime})d\tau^{\prime}.\label{recursion1}
\end{equation}
To our knowledge, this equation for the population growth has not been derived previously, and will allow us to find an analytic solution for the population growth. We find that within our model of correlated generation times, we are able to find an analytic solution. We assume the variance of the noise term $\sigma_{\kappa}^2$ is small and Taylor expand Eq. (\ref{Langevin}) in the main text around $\langle \tau\rangle$ to obtain an approximate auto-regressive process of $\tau$,
\begin{equation}
 \tau^{\prime}=C_{\tau}\tau+ b +\chi\label{tau_approx}.
\end{equation}
Here $b$ is chosen such that Eq. (\ref{tau_approx}) has the same $\langle\tau\rangle$ as that of Eq. (4) in the main text and the variance of the noise $\chi$ is chosen such that Eq. (\ref{tau_approx}) has the same $\sigma_{\tau}^2$ as that of Eq. (4) in the main text. From Eq. (\ref{tau_approx}) it follows that:
\begin{equation}
	h(\tau^{\prime}|\tau)=\frac{1}{\sqrt{2\pi\sigma_{\chi}^2}}\exp\Big(-\frac{(\tau^{\prime}-C_{\tau}\tau-b)^2}{2\sigma_\chi^2}\Big).
\end{equation}
We next introduce an ansatz: $B(\tau)=\exp(-C\tau)$, with an unknown parameter $C$ to be determined. Direct evaluation of the right hand side of Eq. (\ref{recursion1}) leads to
\begin{subequations}
	\begin{align}
		C&=\frac{C_{\tau}\Lambda_p}{1-C_{\tau}}, \\
		\Lambda_p=&\frac{2\ln(2)/\langle \tau \rangle}{1+\sqrt{1-2\ln(2)\sigma_{\tau}^2/\langle \tau \rangle^2 F(C_{\tau})}},\label{lambda_random_generation}
	\end{align}
\end{subequations}
with $F(C_{\tau})=(1+C_{\tau})/(1-C_{\tau})$, $\langle\tau\rangle=b/(1-C_{\tau})$ and $\sigma_{\tau}^2=\sigma_{\chi}^2/(1-C_{\tau}^2)$. Through the derivation, we find that the population growth rate is only determined by the mother-daughter correlation of generation times, namely, $h(\tau^{\prime} |\tau)$ and independent of the sister-sister correlation.  In Appendix D, we discuss in detail the statistics of the generation time distributions on the population tree and the corresponding exact solutions.

We simulate an asynchronous growing population based on Eq. (\ref{Langevin}) in the main text and compute the resulting population growth rate (see numerical details in Appendix A). The numerical results match the theoretical prediction, Eq. (\ref{lambda_random_generation}) well (Fig \ref{figures1}(a)). Numerical tests on $A(\tau)$ are shown in the Fig. \ref{figures1}(b) (see numerical details in Appendix C), which match the theoretical prediction as well. Furthermore, we test the validity of the results of the random-generation-time model to the random-growth-rate model and size-controlled-model in Fig. \ref{figures2}. We find that the random-generation-time model works well for the random-growth-rate model since the generation time is only determined by the single-cell growth rate in this model (Fig. \ref{figures2}(a)). However, the random-generation-time model breaks down for the size-controlled-model (Fig. \ref{figures2}(b))

\section{Appendix C: Numerical simulations of $A(\tau)$}
In the random-generation-time model, we compute the amplitude of the exponential growth $A(\tau)$ by simulating a population starting from a single cell with generation $\tau$. The numerical results, averaged over 30 samples, are shown in Fig. \ref{figures1}(b) (where the ratio between $A(\tau)$ and $A(1/2)$ is shown).

\section{Appendix D: Analytical solution of the tree distribution of generation times}
In Ref. \cite{Lin2017}, it was shown that the tree distribution of generation times $f_0(\tau)$ is distinct from the lineage distribution $f(\tau)$ in the presence of finite correlation between mother and daughter cells. In general, the formula to compute the population growth rate is 
\begin{equation}
2\int_0^{\infty} e^{-\Lambda_p \tau}f_0(\tau)=1.\nonumber
\end{equation}
Even though Powell did not explicitly point out the physical meaning of $f_0(\tau)$, he managed to derive a recursive equation of $f_0(\tau)$ \cite{Powell1956},
\begin{equation}
f_0(\tau)=2\int_0^{\infty} e^{-\Lambda_p \tau^{\prime}} h(\tau |\tau^{\prime})f_0(\tau^{\prime})d\tau^{\prime}\nonumber,
\end{equation}
However, he did not propose analytical results from the recursive equations, or intuition to interpret the finite correlation.
We analytically find that when the lineage distribution $f(\tau)$ is Gaussian, so is the tree distribution of generation times $f_0(\tau)$, with the same variance but with a different mean
\begin{equation}
\langle t\rangle_0=\frac{1+a\sqrt{1-2\ln(2)\frac{1+a}{1-a}\sigma_{\tau}^2}}{1+a},\label{t0}
\end{equation}
and when $a=0$, $\langle t\rangle_0=1$ as expected. We find that a positive $a$ tends to bias the tree distribution towards cells with shorter generation times, consistent with the prediction that a positive correlation between generation times of mother and daughter cells increases the population growth rate.

\section{Appendix E: High-order correction of the population growth rate}
In the main text, we show that the asymptotic formula of population growth rate derived from the random-growth-rate model without time-additive or size-additive noise provides a very good approximation to the situation even with finite time-additive and size-additive noise to the leading order. 

Here we first discuss the higher-order correction due to time-additive noise $\sigma_{\xi}$. The population growth rate is a function of the three variables $\sigma_\lambda$, $\sigma_\xi $ and $\alpha$. It seems plausible, however, that it will be a differentiable function of $\alpha$ and the two variances $\sigma^2_\xi$ and $\sigma^2_\lambda$: for instance, when the growth rate and time-additive noise are normally distributed, it is the variance which enters the formula for the distribution. Under this assumption we have $\Lambda_p = \Lambda_p(\alpha, \sigma^2_\xi, \sigma^2_\lambda)$ and we may proceed to perform a second order Taylor expansion of this multivariate function. Since for $\sigma_{\lambda}=0$ we know that the population growth rate must be strictly equal to the single-cell growth rate and independent of $\alpha$ and $\sigma_\xi$, the lowest order contribution of $\sigma_\xi$ must be of the form $C_2(C_{\lambda},\alpha) \sigma_\xi^2 \sigma_\lambda^2$. Therefore we
have:
\begin{align}
\Lambda_p(\alpha,\sigma_{\lambda}^2,\sigma_{\xi}^2)\approx 1-C_1(C_{\lambda})\sigma_{\lambda}^2+C_2(C_{\lambda},\alpha)\sigma_{\xi}^2\sigma_{\lambda}^2. \label{lambdap}
\end{align}
Here we have set the mean growth rate $\langle\lambda\rangle=1$ for simplicity. The first two terms are the same as Eq. (10) in the main text. $C_{\lambda}$ is the correlation coefficient between the logarithms of growth rates introduced in Eq. (1) in the main text. Because $\sigma_{\xi}$ is on the order $0.1$ in general \cite{Taheri2015}, the high-order correction typically contributes a negligible correction to the second term, which makes it hard to detect numerically. We numerically fit $\Lambda_p$ to the ansatz Eq. (\ref{lambdap}) (Fig. \ref{figures3}(a)), and plot the resulting $C_2$ as function of $C_{\lambda}$ and $\alpha$ (Fig. \ref{figures3}(b)). The same argument also applies to the size-additive noise and we also do the same analysis for size-additive noise with $\sigma_{\xi}$ replaced by $\sigma_{\delta}$ and get a similar result (Fig. \ref{figures3}(c)).
\bibliography{lin}

\end{document}